\documentclass[10pt]{article}

\usepackage{graphicx}
\usepackage[T1]{fontenc}
\usepackage[utf8]{inputenc}
\usepackage{amssymb}
\usepackage{amsfonts}
\usepackage{dsfont}
\usepackage{mathtools}
\usepackage{amsthm}
\usepackage{amsmath}
\usepackage{relsize}
\usepackage{textcomp}
\usepackage{eurosym}
\usepackage{stmaryrd}
\usepackage{xcolor}
\usepackage[multiple]{footmisc}
\usepackage{bigints}
\usepackage{geometry}
\newtheorem{rem}{Remark}

\begin{document}

\title{Market Making in Spot Precious Metals}

\author{Alexander \textsc{Barzykin}\footnote{HSBC, 8 Canada Square, Canary Wharf, London E14 5HQ, United Kingdom, alexander.barzykin@hsbc.com.} \and Philippe \textsc{Bergault}\footnote{Université Paris Dauphine-PSL, Ceremade, Place du Maréchal de Lattre de Tassigny, 75116 Paris, France, bergault@ceremade.dauphine.fr.} \and Olivier \textsc{Guéant}\footnote{Université Paris 1 Panthéon-Sorbonne, UFR 27 Mathématiques et Informatique, Centre d’Economie de la Sorbonne, Paris,
France, olivier.gueant@univ-paris1.fr.}}
\date{}

\maketitle
\setlength\parindent{0pt}

\begin{abstract}

The primary challenge of market making in spot precious metals  is navigating the liquidity that is mainly provided by futures contracts. The Exchange for Physical (EFP) spread, which is the price difference between futures and spot, plays a pivotal role and exhibits multiple modes of relaxation corresponding to the diverse trading horizons of market participants. In this paper, we model the EFP spread using a nested Ornstein-Uhlenbeck process, in the spirit of the two-factor Hull-White model for interest rates. We demonstrate the suitability of the framework for maximizing the expected P\&L of a market maker while minimizing inventory risk across both spot and futures. Using a computationally efficient technique to approximate the solution of the Hamilton-Jacobi-Bellman equation associated with the corresponding stochastic optimal control problem, our methodology facilitates strategy optimization on demand in near real-time, paving the way for advanced algorithmic market making that capitalizes on the co-integration properties intrinsic to the precious metals sector.

\vspace{2mm}

\end{abstract}

\section*{Introduction}

Recent decades have seen a profound transformation in financial markets, driven largely by the widespread adoption of electronification. This digital evolution has been paralleled by the development of new decision-making tools for market makers and the emergence of systematic market making for most asset classes. The case of stocks traded through central limit order books has been discussed extensively (see~\cite{guilbaud2013optimal} for a good example using stochastic optimal control), but OTC markets have also been addressed using the modelling framework introduced by Avellaneda and Stoikov in \cite{avellaneda2008high} (see for instance~\cite{barzykin2023dealing} for the FX market). These market making models have helped dealers to manage inventory risk through strategic hedging and quote skewing. In single-asset models, optimal quotes typically vary monotonically with current inventories, influenced by factors such as risk aversion, clients' pricing sensitivity, and the liquidity on external platforms. Dealers aim to maximize client flow internalization to mitigate the cost of external execution and reduce market impact, hedging only when inventories exceed franchise-dependent thresholds (on the internalization-externalization dilemma, see for instance~\cite{barzykin2022market} and~\cite{butz2019internalisation}).\\

Multi-asset extensions of market making models have been built to help dealers manage their risk at the portfolio level (see for instance~\cite{barzykin2023dealing} and~\cite{evangelista2020closed}). Market makers often deal indeed with large portfolios of assets of very different liquidity. Illiquid assets may be difficult to internalize and costly to execute in the market. However, the risk associated with illiquid assets may sometimes be partially offset by positions in other instruments that are more liquid, leaving time to unwind the illiquid asset with less cost. In FX markets, illiquid instruments can also be traded via more liquid legs, introducing both complexities and opportunities for the dealer (see \cite{barzykin2023dealing} and \cite{cartea2020trading}). Most existing models consider the case of asset price dynamics driven by correlated Brownian motions. \\

Diverging from conventional approaches that focus solely on correlated asset price dynamics, our model capitalizes on the unique benefits of using co-integrated assets for hedging, regardless of whether the market maker proposes quotes in those assets. Co-integration offers indeed a special opportunity for market makers to tap into enhanced liquidity pools and benefit from mean reversion.  This is particularly relevant in the case of a spot dealer hedging their position through futures, as in the case of precious metals market. While the interbank spot market does exist, precious metals futures are considerably more liquid, with tighter spreads and higher volumes. Since the price difference between futures and spot -- so-called Exchange for Physical (EFP) spread -- is primarily driven by swap rates which are relatively stable over the market maker's intraday risk horizon, the co-integration assumption between spot and futures prices is very natural.\footnote{There are other asset classes were the same mathematical ideas could be used. In FX, one can think of market making involving Non-Deliverable Forwards and onshore spot liquidity for instance.}\\

\begin{figure}[h]
\centering
\includegraphics[width=0.79\textwidth]{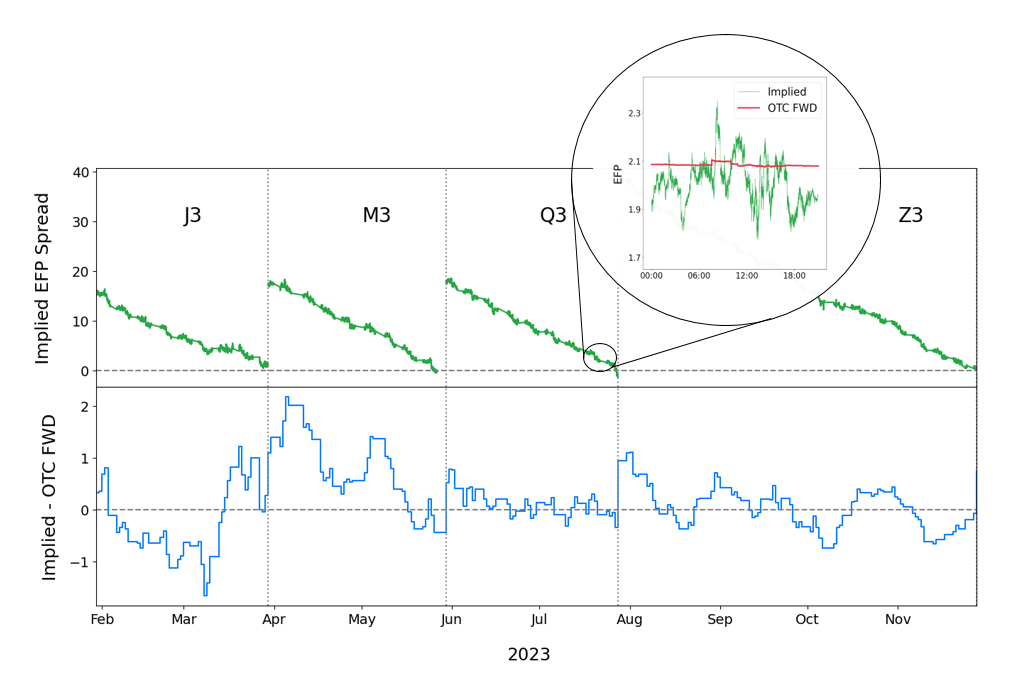}\\
\caption{EFP spreads in basis points implied from four active futures contracts and spot mid prices of gold against USD in 2023 (green).
Zoom in on 21 July also shows OTC forward rate -- OTC FWD -- (red) and demonstrates intraday mean reversion. Daily median difference in basis points between implied EFP and OTC forward rate (blue) illustrates mean reversion on a weekly scale.}
\label{efp_history}
\end{figure}

Our modelling framework builds on the classical foundation laid by~\cite{avellaneda2008high} and~\cite{cartea2014buy} (see~\cite{gueant2016financial} for a detailed discussion). In this framework, a single underlying instrument (spot) price is modelled by an arithmetic Brownian motion. The arrival rates and size distribution of client trades are modelled with predetermined intensity kernels. Going beyond the initial academic literature and building on recent advances, our model considers a dealer quoting in spot who can hedge (with costs) in both spot and futures. The novelty comes from the modelling of EFP by a nested Ornstein-Uhlenbeck (hereafter nested OU) process, as inspired by the observation of multiple relaxation times in the market ranging from hours to days, possibly related to different trading horizons of different types of traders, as illustrated in 
Figure~\ref{efp_history}. This model is in the spirit of the two-factor Hull-White model for interest rates \cite{hull1994numerical}. However, the observed EFP volatility significantly exceeds the underlying interest rate volatility so the mechanism is not purely driven by interest rates but also influenced by the cost of physical delivery, apart from speculation. This article expands, therefore, the now classical stochastic optimal control framework for market making by incorporating the existence of co-integrated and liquid assets. Not only does it address a gap in the existing academic literature on market making models but also provides a new class of price dynamics compatible with the approximation techniques developed in~\cite{evangelista2020closed}.\\

In the forthcoming sections, we begin by introducing our model alongside the key equations that define the optimal strategies for market makers in the spot precious metals market. We also discuss the merits and limitations of utilizing nested OU processes. Subsequently, we demonstrate the applicability of the method from~\cite{evangelista2020closed} to our general framework. We conclude with an in-depth numerical analysis focused on the gold market, illustrating the practical implications of our model.

\section*{The model}

\subsection*{State variables and optimal control problem}

We consider a spot market maker in a given precious metal. We denote by $(S_t)_{t\ge 0}$ a reference spot price for that metal and assume that the market maker continuously streams to clients a pricing ladder at the bid, $S^{b}(t,z) = S_t - \delta^{b}(t,z)$, and at the ask, $S^{a}(t,z) = S_t +\delta^{a}(t,z)$, where $z$ denotes the sizes in ounces (oz). Transaction likelihood is assumed to depend solely on the distance between the proposed prices and the reference spot price. Subsequently, we introduce two intensity functions for bid and ask: $(z, \delta) \mapsto \Lambda^{b}(z, \delta)$ and $(z, \delta) \mapsto \Lambda^{a}(z, \delta)$. We assume that the functions $\Lambda^{b}$ and $\Lambda^{a}$ take the form\footnote{Generalizations do not raise difficulties.}
$$\Lambda^{b}(z,\delta)=\Lambda^{a}(z,\delta) = \Lambda (z,\delta) = \lambda(z) f(\delta) \quad \text{with} \quad f(\delta) =  \frac{1}{1+e^{\alpha + \beta \delta}},\quad  \beta>0 \; \text{and}\; \alpha \in \mathbb R.$$

The market maker is also capable of hedging their position by trading on various platforms, using either or both spot and futures. The execution rates associated with the market maker's activities are modeled by the processes $(v^S_t)_{t\ge 0}$ and $(v^F_t)_{t\ge 0}$ for spot and futures, respectively.\\

We denote by $(F_t)_{t\ge 0}$ the price process of the futures contract and by $(E_t)_{t\ge 0}$ the EFP process, i.e. for all $t \ge 0$, $F_t = S_t + E_t$.\footnote{We consider de-seasonalized futures prices with interest rate dependence on time to maturity factored out because interest rates have very low intraday volatility (see Figure~\ref{efp_history}).}
 In what follows, we consider a Brownian dynamics for $(S_t)_{t\ge 0}$ and a nested OU dynamics\footnote{It is possible to expand the nested dynamics to include more than  one level if there is statistical or practical need.} for $(E_t)_{t\ge 0}$, i.e.,
$$dS_t = \sigma_S dW^S_t,  \qquad \sigma_S > 0$$ and
$$dE_t = -k_E \left(E_t - D_t \right) dt + \sigma_E dW^E_t, \qquad k_E, \sigma_E > 0,$$
with
$$dD_t = -k_D \left(D_t - \bar{D}\right)dt + \sigma_D dW^D_t, \qquad k_D, \sigma_D \ge 0, \quad \bar{D} \in \mathbb{R},$$
where $(W_t^S, W_t^E, W_t^D)_{t\ge 0}$ is a three-dimensional Brownian motion with correlation matrix $R$. In what follows, we denote by $\Sigma$ the variance-covariance matrix $\text{diag}(\sigma_S, \sigma_E, \sigma_D)\; R\; \text{diag}(\sigma_S, \sigma_E, \sigma_D)$.\\

We denote the inventory processes of the market maker by $(q^S_t)_{t\ge 0}$ and $(q^F_t)_{t\ge 0}$ for spot and futures contract respectively. Mathematically, the dynamics of $(q^S_t)_{t\ge 0}$ is formalized by considering the random measures $J^{b}(dt,dz)$ and $J^{a}(dt,dz)$ modeling the times and sizes of OTC trades on the bid and ask sides, respectively. Subsequently, the inventory dynamics for spot can be expressed as
$$dq^S_t = \int\limits_{z=0}^{\infty} zJ^{b}(dt,dz) - \int\limits_{z=0}^{\infty} zJ^{a}(dt,dz) + v^S_t dt,$$
while the dynamics for the inventory in the futures contract is modelled by
$$dq^F_t = v^F_t dt.$$

The resulting cash process $(X_t)_{t\ge 0}$ of the market maker writes 
$$dX_t =  \int\limits_{z=0}^{\infty} S^{a}(t,z)zJ^{a}(dt,dz) -  \int\limits_{z=0}^{\infty} S^{b}(t,z)zJ^{b}(dt,dz) - v^S_tS_t dt - L^S(v^S_t) dt - v^F_t F_t dt - L^F(v^F_t) dt,$$
where the terms $L^S(v^S_t)$ and  $L^F(v^F_t)$ account for spread costs along with the temporary price impact of the market maker upon externalizing.\footnote{We have not considered permanent market impact here for the sake of simplicity. We might assume that both spot and futures external transactions (i.e. the processes $(v^S_t)_{t\ge 0}$ and $(v^F_t)_{t\ge 0}$) affect the spot price equally, without altering the EFP spread. Under the assumption of linear permanent market impact, our model remains effective. For more details on market impact, see the recent paper \cite{hey2023cost} on the importance to have a good market impact model.} The functions $L^S$ and $L^F$ are typically nonnegative, strictly convex, and asymptotically super-linear. Here, we consider $$L^S(v) = \psi^S |v| + \eta^S v^2 \qquad \text{and} \qquad L^F(v) = \psi^F |v| + \eta^F v^2.$$

The market maker wants to maximize the expected utility of the Mark-to-Market value of their portfolio at the end of the period $[0,T]$ minus a penalty corresponding to remaining inventories. We assume that the market maker has a CARA utility function, and maximizes
\begin{equation*}
\mathbb{E} \left[-\exp\left(-\gamma \left(X_T + q^S_T S_T + q^F_T F_T - K^S (q^S_T)^2 - K^F (q^F_T)^2 \right)\right) \right]
\end{equation*}
by selecting two $\mathcal{P} \otimes \mathcal{B}(\mathbb{R}_{+}^{*})$-measurable processes\footnote{Here $\mathcal{P}$ denotes the $\sigma$-algebra of $\mathbb{F}$-predictable subsets of $\Omega \times[0,T]$ and $\mathcal{B}(\mathbb{R}_{+}^{*})$ denotes the Borelian sets of $\mathbb{R}_{+}^{*}$.} $\delta^{b}$, $\delta^{a}$ bounded from below and two $\mathcal{P}$-measurable processes $v^S$, $v^F$, with
$$\mathbb E \left[ \int_0^T \left(v^S_t\right)^2 d t \right] < +\infty \qquad \text{and} \qquad \mathbb E \left[ \int_0^T \left(v^F_t\right)^2 d t \right] < +\infty,$$
and where $\gamma$ indicates the market maker's risk aversion and $K^S, K^F \ge 0$ are the penalty coefficients.

\subsection*{Solution}

We denote by $u:[0,T]\times \mathbb R^6\rightarrow \mathbb{R}$ the value function of this stochastic control problem. The Hamilton-Jacobi-Bellman equation associated with it is:
\begin{eqnarray}
0 &=& \partial_t u - k_E  \left(E  - D \right)\partial_{E}u - k_D \left(D-\bar D \right)\partial_D u+ \frac 12 \text{Tr}(\Sigma \nabla^2_{SED}u) + \mathcal L^b u + \mathcal L^a u \nonumber \\
&& +\ \underset{v^S}{\sup} \left( v^S\partial_{q^S} u - \left( L^S(v^S) + v^SS \right) \partial_x u\right) + \underset{v^F}{\sup} \left( v^F\partial_{q^F} u- \left( L^F(v^F) + v^F(S+E) \right) \partial_x u\right),\label{HJBu}
\end{eqnarray}
with terminal condition $u(T,x, q^S, q^F,S, E, D) = -\exp\left(-\gamma \left(x + q^S S + q^F (S+E) - K^S (q^S)^2 - K^F (q^F)^2 \right)\right)$, where
$$\mathcal{L}^b u (t,x,q^S\!,q^F\!,S,E,D)\! =\! \int\limits_{0}^{\infty}\! \underset{\delta^b}{\sup}  f(\delta^b) \left( u(t,x-z(S-\delta^b), q^S\!+z, q^F\!, S, E, D) - u(t,x,q^S\!,q^F\!,S,E,D) \right) \lambda(z) \, dz$$
$$\mathcal L^a u (t,x,q^S\!,q^F\!,S,E,D)\! =\!  \int\limits_{0}^{\infty}\! \underset{\delta^a}{\sup} f(\delta^a) \left( u(t,x+z(S+\delta^a), q^S\!-z, q^F\!, S, E, D) - u(t,x,q^S\!,q^F\!,S,E,D) \right) \lambda(z) dz.$$
This equation can be simplified through the use of the ansatz
$$u(t,x,q^S,q^F,S,E,D) = -\exp\left(-\gamma \left( x+q^S S + q^F(S+E) + \theta(t,q^S,q^F,E,D) \right)\right),$$
where $\theta:[0,T]\times \mathbb R^4\rightarrow \mathbb{R}$ is a differentiable function such that $\theta(T,q^S,q^F,E,D) = - K^S (q^S)^2 - K^F (q^F)^2$.\\

 The partial differential equation associated with $\theta$ is indeed the following:
\begin{eqnarray}
0 &=& \partial_t \theta - k_E  \left(E  - D \right)\left(q^F + \partial_{E}\theta\right)  - k_D \left(D-\bar D \right)\partial_D \theta+\frac 12 \text{Tr}(\widetilde \Sigma\nabla^2_{ED}\theta) \nonumber\\ && - \frac{\gamma}{2}\begin{pmatrix} q^S + q^F \\ q^F + \partial_E \theta \\ \partial_D \theta \end{pmatrix}^\intercal \Sigma \begin{pmatrix} q^S + q^F \\ q^F + \partial_E \theta \\ \partial_D \theta \end{pmatrix}+ \mathcal J_H \theta + \mathcal H^S \left(\partial_{q^S}\theta\right) +  \mathcal H^F \left(\partial_{q^F}\theta\right) \label{HJB}
\end{eqnarray}
where $\widetilde \Sigma$ is the submatrix of $\Sigma$ obtained by removing the first row and the first column, $\mathcal H^S$ and $\mathcal H^F$ are the Hamiltonian functions defined by
$$\mathcal H^S:p\in\mathbb{R} \ \mapsto \underset{v^S }{\sup} \left(v^Sp- L^S(v^S)   \right) \quad \text{and} \quad\mathcal H^F:p\in\mathbb{R} \ \mapsto \underset{v^F }{\sup} \left(v^Fp- L^F(v^F)   \right),$$ and
\begin{eqnarray*}\mathcal J_H \theta(t,q^S, q^F, E , D) =  \int\limits_{0}^{\infty} zH \left(z, \mathcal J^+ \theta (t,q^S, q^F, E , D, z)\right)\lambda(z) dz + \int\limits_{0}^{\infty} zH \left(z, \mathcal J^- \theta (t,q^S, q^F, E , D, z)\right)\lambda(z) dz
\end{eqnarray*}
with $H:(z,p)\in (0, +\infty) \times \mathbb{R} \mapsto \underset{\delta }{\sup} \frac{f(\delta)}{\gamma z}(1-e^{-\gamma z (\delta-p)})$ and
$$\mathcal J^\pm \theta (t,q^S, q^F, E , D, z) = \frac{\theta(t,q^S, q^F, E , D) -  \theta(t,q^S\pm z, q^F, E , D) }{z}.$$

Under classical assumptions on the intensities (here on the function $f$ -- see for instance \cite{gueant2016financial}), it can be proved that, given a smooth solution to the equation \eqref{HJB}, the optimal controls are given by\footnote{Because $q^S$ is not a continuous process, we consider in \eqref{eq:strat_system} the left limit of $q^S$ at time $t$ denoted by $q^S_{t-}$.}
\begin{equation}
\left\{
\begin{aligned}
\delta^{b*}(t,z) &= \bar\delta \left(z, \mathcal J^+ \theta(t,q^S_{t-}, q^F_t, E _t, D_t,z)  \right) \\
\delta^{a*}(t,z) &= \bar \delta \left(z, \mathcal J^- \theta(t,q^S_{t-}, q^F_t, E _t, D_t,z) \right) \\
v^{S*}_t &= \left({\mathcal H^S}\right)'\left( \partial_{q^S}\theta(t,q^S_{t-}, q^F_t, E _t, D_t) \right) \\
v^{F*}_t &= \left({\mathcal H^F}\right)'\left( \partial_{q^F}\theta(t,q^S_{t-}, q^F_t, E _t, D_t) \right)
\end{aligned}
\right.
\label{eq:strat_system}
\end{equation}
where $\bar \delta(z, p) = f^{-1} \left(\gamma z H(z,p)-\partial_p H (z,p)  \right)$.

\subsection*{Remarks on nested OU processes}

Optimal quoting and hedging strategies in our model are contingent upon current inventories and the current values of the processes $(E_t)_{t\ge 0}$ and $(D_t)_{t\ge 0}$. While the EFP and inventories are directly observable, $(D_t)_{t\ge 0}$ is not, introducing a challenge for practical implementation.\\

Assuming $(D_t)_{t\ge 0}$ is fixed as in a simple OU model where $k_D = \sigma_D = 0$ leads to market makers engaging in overly confident statistical arbitrage. This could lead indeed to strategies that are strongly reliant on the EFP mean reverting to the constant value of the process $(D_t)_{t\ge 0}$. To encourage caution, our model introduces variability in $(D_t)_{t\ge 0}$ by setting $k_D$ and $\sigma_D$ to positive values. These parameters can then be viewed as hyperparameters, allowing market makers to modulate their confidence. With that vision, the current value of the process $(D_t)_{t\ge 0}$ might be estimated or pragmatically chosen.\\

Another (more rigorous) point of view consists in statistically estimating all the
parameters and filtering the signal to get an approximation of the process
$(D_t)_{t\ge 0}$ at all times. For that purpose, we can assume a correlation
structure $R = \begin{pmatrix} 1 & \rho & 0 \\ \rho & 1 & 0\\ 0 & 0 &
1 \end{pmatrix}$ with $\rho \in [-1,1]$. Then, all the parameters can
be estimated using classical maximum likelihood (without observing
$(D_t)_{t\ge 0}$) methods because $(E_t)_{t\ge 0}$ is a Gaussian process whose
covariance function is known in closed form. Once parameters are estimated, we use filtering techniques to write
\begin{align*}
\begin{cases}
    dS_t&= \sigma_S d\widehat{W}_t^S\\
    dE_t &= -k_E \left(E_t - \widehat{D}_t \right) dt + \sigma_E d\widehat{W}^E_t\\
    d\widehat{D}_t &= -k_D \left(\widehat{D}_t - \bar{D}\right)dt + \frac{1}{\sqrt{1-\rho^2}}\frac{k_E}{\sigma_E} \nu_t^2 d\widehat{W}^D_t
\end{cases}
\end{align*}
where
$$\widehat{D}_t = \mathbb{E}\left[D_t | (S_s)_{s\le t}, (E_s)_{s\le t} \right], \quad \nu^2_t = \mathbb{V}\left(D_t | (S_s)_{s\le t}, (E_s)_{s\le t}\right),$$
and $$\widehat{W}_t^S = W_t^S, \quad \widehat{W}^E_t = W^E_t + \frac{k_E}{\sigma_E}\int_0^t (D_s-\widehat{D}_s) ds, \quad \widehat{W}^D_t =  \frac{\widehat{W}^E_t - \rho \widehat W_t^S}{\sqrt{1-\rho^2}}$$ define a three-dimensional Brownian motion adapted to the natural filtration of the processes $(S_t)_{t\ge 0}$ and $(E_t)_{t\ge 0}$ with correlation structure $\widehat{R} = \begin{pmatrix} 1 & \rho & 0 \\ \rho & 1 & \sqrt{1-\rho^2} \\ 0 & \sqrt{1-\rho^2} & 1 \end{pmatrix}$ of rank $2$.\\

Using standard Bayesian filtering techniques, $(\nu_t^2)_{t\ge 0}$ is in fact deterministic and satisfies
$$\frac{d\nu_t^2}{dt} = - \frac{1}{1-\rho^2}\frac{k_E^2}{\sigma_E^2} \nu_t^4 - 2 k_D \nu_t^2 + \sigma_D^2.$$
In particular, assuming we have observed the spot and futures prices for a long time, we can replace $\nu_t^2$ by its asymptotic value:
$$\nu_\infty^2 = \frac{\sigma_D^2}{k_D + \sqrt{k_D^2 + \frac{1}{1-\rho^2}\frac{k_E^2}{\sigma_E^2}\sigma_D^2 }}.$$

Our market making problem can then be solved as if $(D_t)_{t \ge 0}$ was observed by replacing the unobservable process $(D_t)_{t\ge 0}$ by its observable counterpart $(\widehat D_t)_{t\ge 0}$, up to the replacement\footnote{When we filter, we remain here in the same family of processes and therefore the Hamilton-Jacobi-Bellman equation remains the same up to the value of those coefficients.} in the partial differential equation \eqref{HJB} of $R$ by $\widehat R$ and $\sigma_D$ by $$\widehat \sigma_D = \frac{1}{\sqrt{1-\rho^2}}\frac{k_E}{\sigma_E} \nu_\infty^2 = \sigma_D \frac{\xi}{k_D + \sqrt{k_D^2 + \xi^2 }}, \quad \text{where} \quad \xi = \frac{1}{\sqrt{1-\rho^2}}k_E\frac{\sigma_D}{\sigma_E}.$$

\section*{Approximation technique}

Approximating numerically the solution $\theta$ of equation \eqref{HJB} using grid methods poses challenges due to the high dimensionality of the state space and the complex geometry of the frequently visited inventory states at optimality. To avoid grids, we adopt a methodology similar to that described in \cite{evangelista2020closed}, which involves approximating equation \eqref{HJB} with a closely related equation. The solution to that equation, denoted by $\check{\theta}$, will be a quadratic polynomial and serve as an approximation to the original value function $\theta$. Consequently, $\check{\theta}$ will be utilized in lieu of $\theta$ in \eqref{eq:strat_system} to derive strategies that are nearly optimal.\\

To obtain our new equation, we employ instead of $L^S$ and $L^F$ the quadratic approximations $\check{L}^S(v) = \eta^S v^2$ and $\check{L}^F(v) = \eta^F v^2$. Subsequently, we substitute $\mathcal{H}^S$ and $\mathcal{H}^F$ in \eqref{HJB} by
$$\check {\mathcal H}^S:p\in\mathbb{R} \ \mapsto \underset{v^S }{\sup} \left(v^Sp- \check L^S(v^S)   \right) = \frac{p^2}{4\eta^S} \qquad \text{and} \qquad \check {\mathcal H}^F:p\in\mathbb{R} \ \mapsto \underset{v^F }{\sup} \left(v^Fp- \check L^F(v^F)   \right) = \frac{p^2}{4\eta^F}.$$
Also, following~\cite{evangelista2020closed}, we approximate the function $H$ by a quadratic function
$$\check H(p) = \alpha_0 + \alpha_1 p + \frac 12 \alpha_2 p^2.$$

We then obtain the following partial differential equation:
\begin{eqnarray}
0 &=& \partial_t \check \theta - k_E  \left(E  - D \right)\left(q^F + \partial_{E}\check \theta\right)  - k_D \left(D-\bar D \right)\partial_D \check \theta+\frac 12 \text{Tr}(\widetilde \Sigma\nabla^2_{ED}\check \theta) \nonumber\\ && - \frac{\gamma}{2}\begin{pmatrix} q^S + q^F \\ q^F + \partial_E \check \theta \\ \partial_D \check \theta \end{pmatrix}^\intercal \Sigma \begin{pmatrix} q^S + q^F \\ q^F + \partial_E \check \theta \\ \partial_D \check \theta \end{pmatrix}+ \mathcal J_{\check H} \check \theta + \frac{1}{4 \eta^S}\left(\partial_{q^S}\check{\theta} \right)^2 +  \frac{1}{4 \eta^F}\left(\partial_{q^F}\check{\theta}\right)^2 \label{HJBquad}
\end{eqnarray}
with terminal condition $\check \theta(T,q^S,q^F,E,D) = - K^S (q^S)^2 - K^F (q^F)^2 $, whose solution $\check \theta$ will be our approximation of $\theta$.\\

As announced above, the interest of the above approximation is that the solution to \eqref{HJBquad} is a polynomial of degree $2$ in $q^S$, $q^F$, $E$, and $D$. Let us make indeed the following ansatz:
$$\check \theta(t,q^S, q^F, E , D) = - \begin{pmatrix}
   q^S\\ q^F\\ E  \\ D 
\end{pmatrix}^\intercal A(t) \begin{pmatrix}
   q^S\\ q^F\\ E  \\ D 
\end{pmatrix} - \begin{pmatrix}
   q^S\\ q^F\\ E  \\ D 
\end{pmatrix}^\intercal B(t) - C(t),$$
where $A:[0,T] \mapsto \mathcal S_4(\mathbb R)$, $B:[0,T] \mapsto \mathbb R^4$, and $C:[0,T] \mapsto \mathbb R$ are differentiable functions such that
$$A(T) =  \begin{pmatrix}
    -K^S  & 0 & 0 & 0 \\
    0 & -K^F & 0 & 0 \\
    0 & 0 & 0 & 0 \\
    0 & 0 & 0 & 0 \\
\end{pmatrix}, \quad  B(T) = \begin{pmatrix}
0\\
0\\
0\\
0
\end{pmatrix} \quad \text{and} \quad C(T) = 0.$$ Plugging this polynomial ansatz into \eqref{HJBquad} yields a system of ODEs for $A$, $B$, and $C$:\footnote{We only report here the equations for $A$ and $B$, as $C$ is irrelevant for the computation of the optimal strategy.}

\begin{align*}
\begin{cases}
    A'(t) &= A(t)M^A A(t) + A(t)U^A + {U^A}^\intercal A(t) +R^A\\
    B'(t) &=  A(t) M^A B(t) +A(t) V^B + {U^A}^\intercal B(t)
\end{cases}
\end{align*}
where
$$M^A = \left(\begin{array}{c|c}
\begin{matrix}
4\alpha_2 \int_0^{+\infty} z\lambda(z) dz + \frac 1{\eta^S} & 0 \\
0 & \frac 1{\eta^F} \\
\end{matrix} & 0_{2\times2} \\
\hline
0_{2\times2} & -2\gamma \widetilde \Sigma \\
\end{array}\right), \quad U^A = 
\left(\begin{array}{c|c}
0_{2\times 2} & 0_{2\times2} \\
\hline
\gamma \begin{pmatrix}
0 & 1 & 0 \\
0 & 0 & 1 \\
\end{pmatrix} \Sigma \begin{pmatrix}
1 & 1 \\
0 & 1 \\
0 & 0\\
\end{pmatrix} & \begin{matrix}
k_E & -k_E \\
0 & k_D \\
\end{matrix}\\
\end{array}\right),$$
$$R^A = -\frac 12 \gamma \left(\begin{array}{c|c}
\begin{pmatrix}
1 & 0 & 0 \\
1 & 1 & 0\\
\end{pmatrix}\Sigma \begin{pmatrix}
1 & 1 \\
0 & 1 \\
0 & 0\\
\end{pmatrix} & 0_{2\times2} \\
\hline
0_{2\times2} & 0_{2\times2}\\
\end{array}\right) - \frac 12 k_E \begin{pmatrix}
    0  & 0 & 0 & 0 \\
    0 & 0 & 1 & -1 \\
    0 & 1 & 0 & 0 \\
    0 & -1 & 0 & 0 \\
\end{pmatrix} \text{\ and \ } V^B = 
\begin{pmatrix}
0\\
0\\
0\\
-2k_D \bar D    
\end{pmatrix}.
$$

This system of ODEs can easily be solved numerically to finally get $\check \theta$. Then, replacing $\theta$ by $\check \theta$ in \eqref{eq:strat_system} leads to the following approximations of the optimal controls:
\begin{equation}
\left\{
\begin{aligned}
\check \delta^{b*}(t,z) &= \bar\delta \left( z(e^S)^\intercal A(t) e^S + 2 \begin{pmatrix}
   q^S\\ q^F\\ E  \\ D 
\end{pmatrix}^\intercal A(t) e^S + (e^S)^\intercal B(t) \right) \\
\check \delta^{a*}(t,z) &= \bar \delta \left( z(e^S)^\intercal A(t) e^S - 2 \begin{pmatrix}
   q^S\\ q^F\\ E  \\ D 
\end{pmatrix}^\intercal A(t) e^S - (e^S)^\intercal B(t) \right) \\
\check v^{S*}_t &= \left(\mathcal H^S\right)'\left( -2(e^S)^\intercal A(t) \begin{pmatrix}
   q^S\\ q^F\\ E  \\ D 
\end{pmatrix} - (e^S)^\intercal B(t)  \right) \\
\check v^{F*}_t &= \left(\mathcal H^F\right)'\left(  -2(e^F)^\intercal A(t) \begin{pmatrix}
   q^S\\ q^F\\ E  \\ D 
\end{pmatrix} - (e^F)^\intercal B(t) \right),
\end{aligned}
\right.
\label{eq:strat_system2}
\end{equation}
where $e^S = \begin{pmatrix}
    1\\ 0\\ 0\\ 0
\end{pmatrix}$ and $e^F = \begin{pmatrix}
    0\\ 1\\ 0\\ 0
\end{pmatrix}$. 

\begin{rem}
It is important to note that in \eqref{eq:strat_system2}, we employ the original functions $\bar{\delta}$, $\mathcal{H}^S$ and $\mathcal{H}^F$ without resorting to quadratic approximations. The only function approximated is $\theta$, which is replaced by $\check{\theta}$. In the terminology of reinforcement learning, this is equivalent to approximating the true value function and then selecting greedy actions based on this approximation. In particular, the spread effects introduced by the functions $L^S$ and $L^F$ are duly considered.
\end{rem}

\section*{Numerical results and discussion}

For illustration, we consider market making in spot gold (XAUUSD) with access to futures liquidity.\\

A typical pricing ladder size discretization is set as $100$, $200$, $500$, $1000$, $2000$ and $5000$ oz. We assume a standard client trade intensity represented by $$\lambda(z) = \begin{cases} 
1600 & \text{if } z = 100 \\
600 & \text{if } z = 200 \\
1000 & \text{if } z = 500 \\
600 & \text{if } z = 1000 \\
120 & \text{if } z = 2000 \\
80 & \text{if } z = 5000
\end{cases} \quad \text{and} \quad f(\delta) = \frac 1{1+ e^{\alpha + \beta\delta}}$$ where price sensitivity parameters are taken to be $\alpha = -0.8$ and $\beta = 5\,\text{bp}^{-1}$. Gold spot and EFP volatility parameters are set to $\sigma_\text{S} = 140\,\text{bp} \cdot \text{day}^{-1/2}$ and $\sigma_E  = 5\, \text{bp} \cdot \text{day}^{-1/2}$, respectively. Spot-EFP correlation is typically small and assumed to be zero in this example. EFP price relaxation rate is $k_E  = 8\,\text{day}^{-1}$. We assume first, as a benchmark, a simple OU framework with $k_D = \sigma_D = 0$ and $\bar{D}$ is fixed at zero. Standard functional dependence of instantaneous market impact on execution rate is assumed: $L^S(v) = \psi^S |v| + \eta^S v^2$ and $L^F(v) = \psi^F |v| + \eta^F v^2$ with
$\psi^S = 0.4$~bp, $\psi^F = 0.2$ bp, $\eta^S = 7\cdot10^{-8}\,\text{bp} \cdot \text{day} \cdot \text{oz}^{-1}$ and 
$\eta^F = 3\cdot10^{-8}\, \text{bp} \cdot \text{day} \cdot \text{oz}^{-1}$.\footnote{The parameters have been selected by analyzing a subset of HSBC market making franchise. However, they should not be considered as representative of HSBC but rather of a typical institutional spot gold dealer. Daily turnover in this setup is approximately \$1B.
Basis point convention is used for price changes, which is an approximation insignificant in practical terms given the time frame of the problem (see \cite{barzykin2022market}).}
We consider a time horizon of $T=1$ hour that ensures convergence towards stationary quotes and hedging rates at time $t=0$. Terminal penalty coefficients are set to zero unless specified otherwise.\\

\begin{figure}[h!]
\centering
\includegraphics[width=0.49\textwidth]{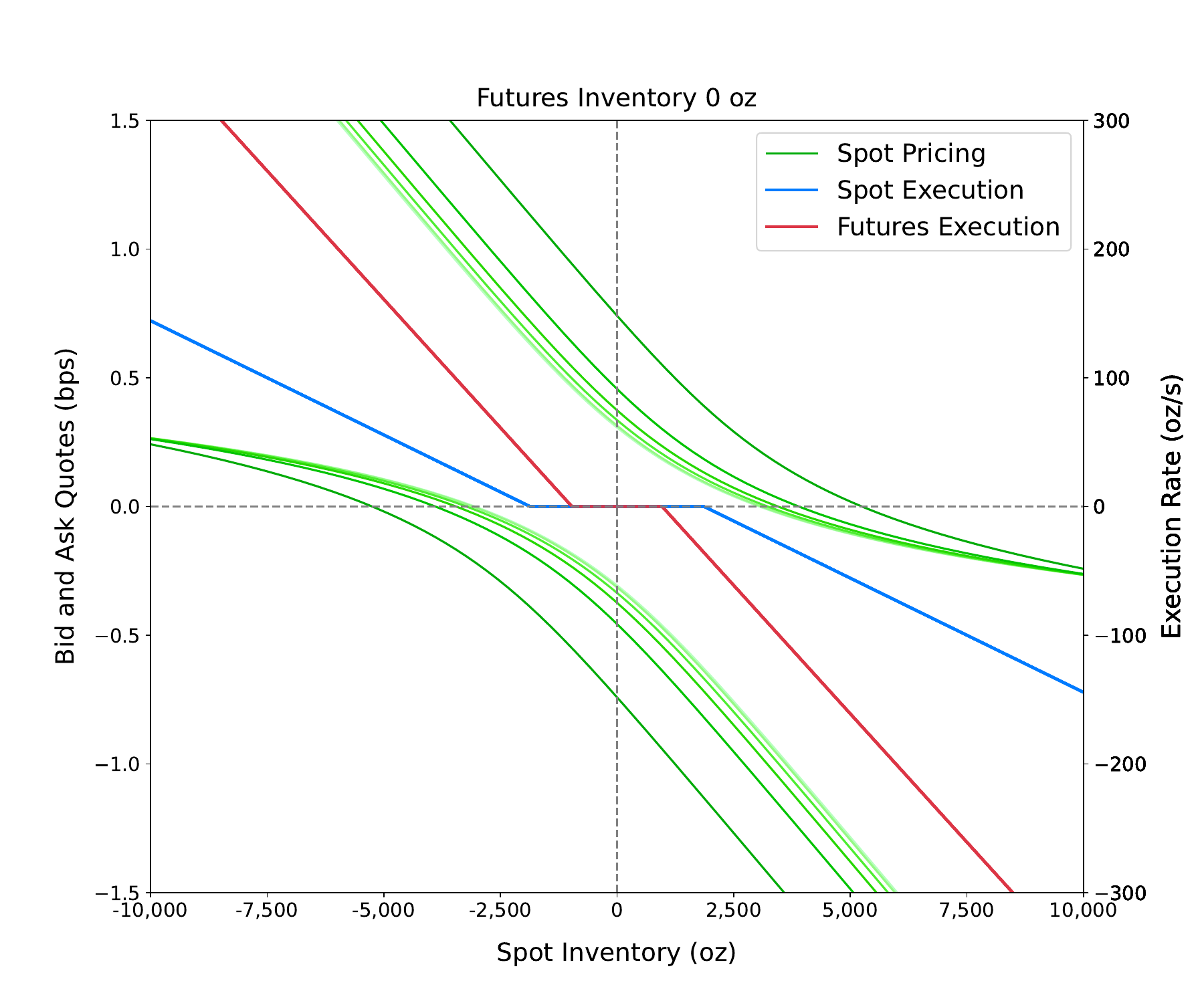}
\includegraphics[width=0.49\textwidth]{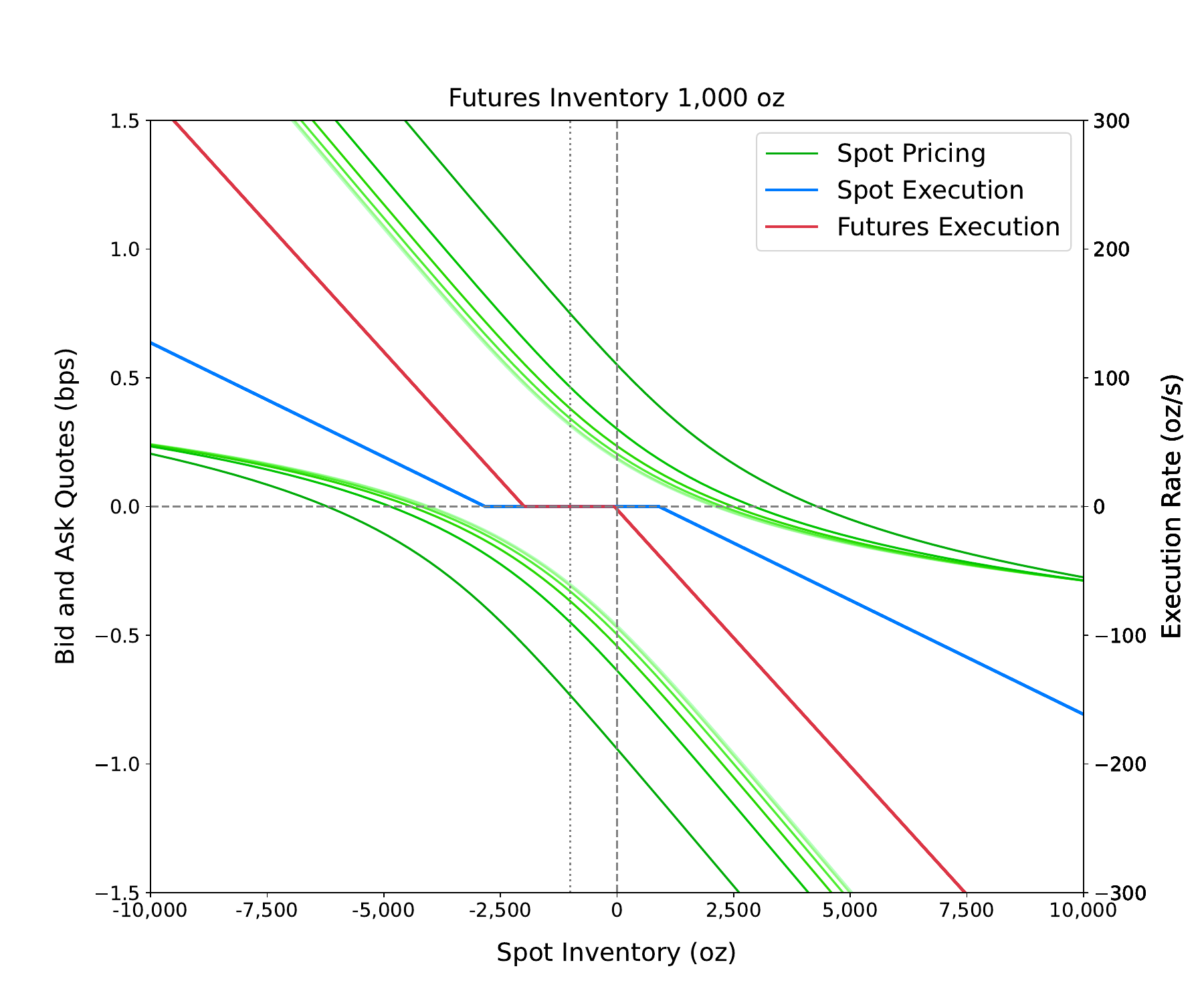}\\
\caption{Optimal gold spot pricing ladder (green), spot execution rate (blue) and futures execution rate (red) as functions of spot inventory
for zero (left) and 1000 oz (right) futures inventory.
EFP price deviation is zero, risk aversion $\gamma = 3\cdot 10^{-4}$, other parameters in the text. Different shades of green correspond to different sizes in the price ladder (lighter colors -- smaller size -- tighter spread). 
}
\label{pricing_hedging}
\end{figure}

Figure~\ref{pricing_hedging} demonstrates the variation of optimal market making controls with spot inventory. When there are no futures in the book, the equilibrium spot position is zero, and the dealer will skew quotes to attract client flow in the direction towards this equilibrium. As noticed in other papers dealing with the internalization-externalization dilemma (see for instance \cite{barzykin2022market}), there exists a pure internalization area where skewing will be considered as the only risk reducing option. Larger positions will, however, also involve hedging in the market. We can see that the onset for hedging with futures is earlier than for spot and the corresponding rate of execution is also higher for futures than for spot.
This is totally understandable given the liquidity (and thus the cost of trading) difference. A non-zero position in futures leads to a shifted quasi-equilibrium in spot, corresponding to an EFP position (where each futures position is paired with another in spot in the opposite direction).\\
\newpage

\begin{figure}[!h]
\centering
\includegraphics[width=0.84\textwidth]{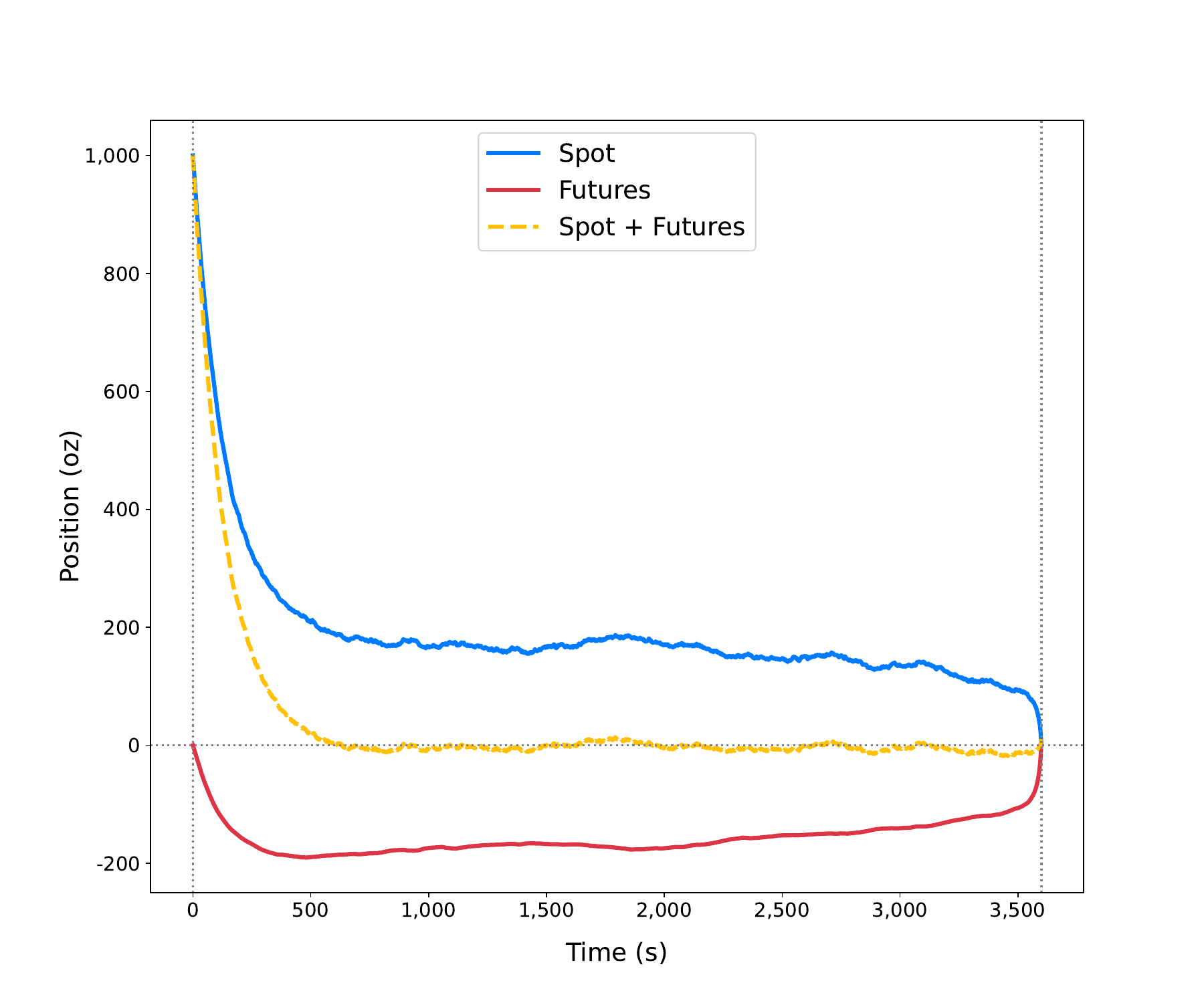}\\
\caption{Average gold spot (blue) and futures (red) inventory relaxation following a 1000 oz client spot trade with a termination condition in 1 hour obtained by numerically averaging $2\cdot 10^4$ Monte Carlo trajectories.
The dashed line (yellow) shows the dynamics of $q^S + q^F$.
}
\label{relaxation}
\end{figure}
\begin{figure}[h!]
\centering
\includegraphics[width=0.84\textwidth]{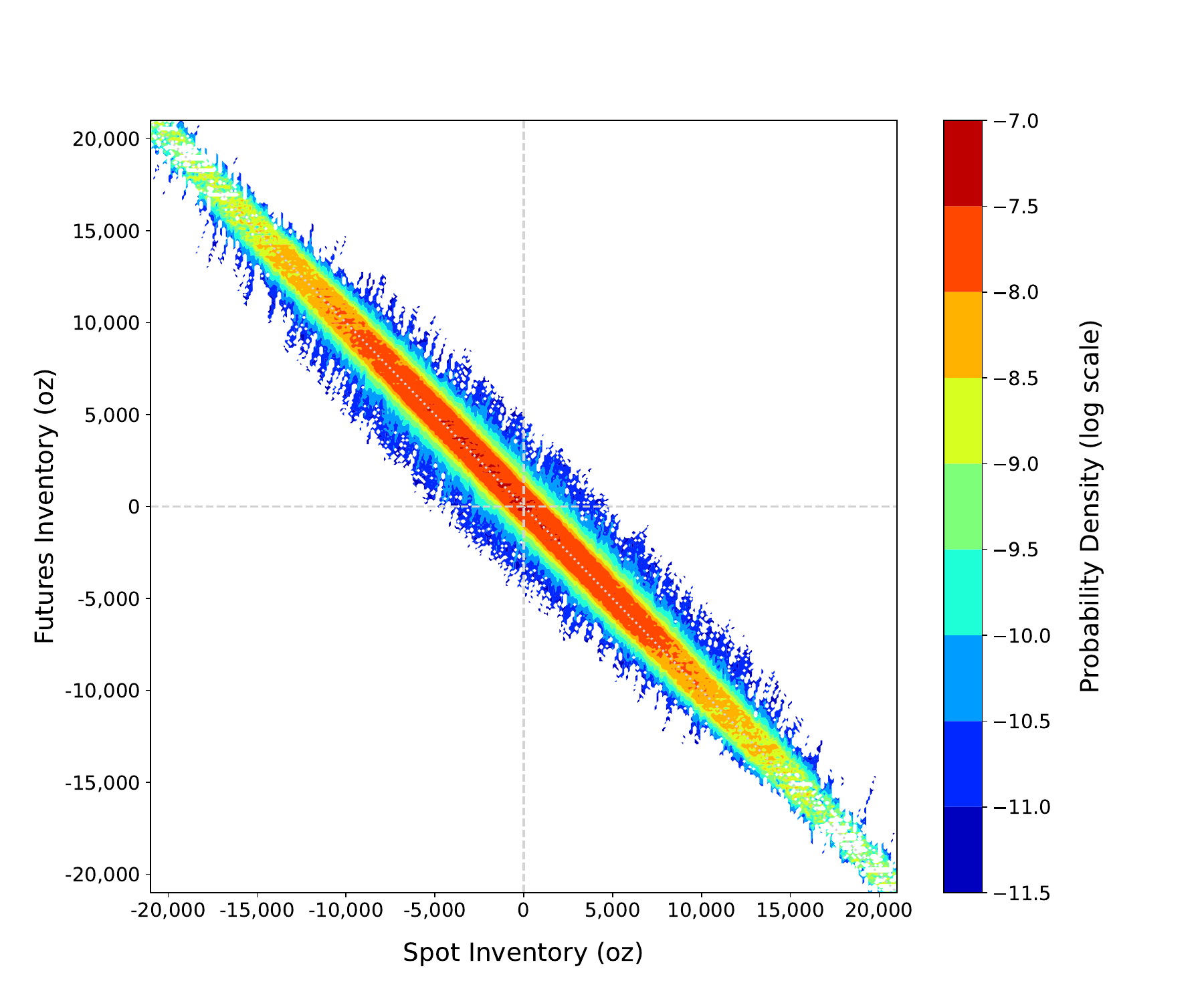}\\
\caption{Inventory probability distribution of a spot gold market maker with access to futures hedging.
}
\label{inventory_pdf}
\end{figure}

Figure~\ref{relaxation} illustrates the expected inventory relaxation following a relatively large 1000 oz OTC trade for  $\gamma = 3\cdot 10^{-4}$. It was calculated via Monte Carlo simulation ($2\cdot 10^4$ trajectories) of the model given the initial condition and optimal controls with terminal penalty\footnote{The effect of terminal penalty is seen only near $T$, so the choice of $K^S$ and $K^F$ is not very important. However, this may change for less liquid markets.} $K^S = K^F = 10^{-3}$. We see that part of the risk is hedged with futures very quickly. The remaining excess spot risk is unwound via skew (attracting the offsetting client flow) and spot execution. Figure~\ref{pricing_hedging} shows that skew can be quite aggressive for this level of risk aversion, with offers better than mid price. Effectively, during this fast relaxation stage, the dealer creates an EFP position which is later unwound very slowly. Figure~\ref{inventory_pdf} shows an inventory probability distribution extracted from a sufficiently long Monte Carlo trajectory ($10^7$ seconds).\footnote{Since the effect of terminal penalty is rather local, this is practically equivalent to carrying simulation over multiple days.} As expected, the system spends most of the time along the low-risk EFP diagonal.\\

\begin{figure}[h!]
\centering
\includegraphics[width=0.49\textwidth]{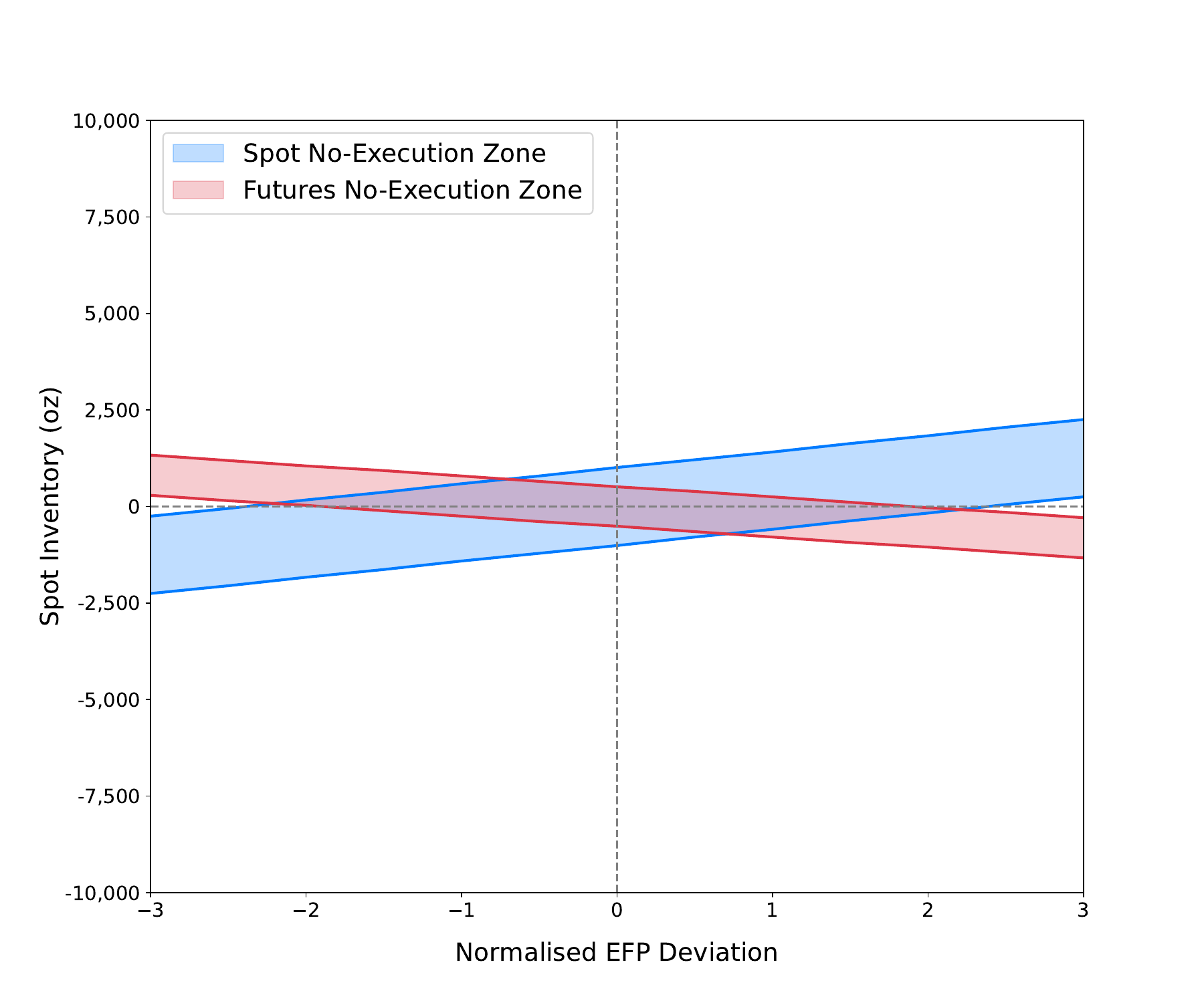}
\includegraphics[width=0.49\textwidth]{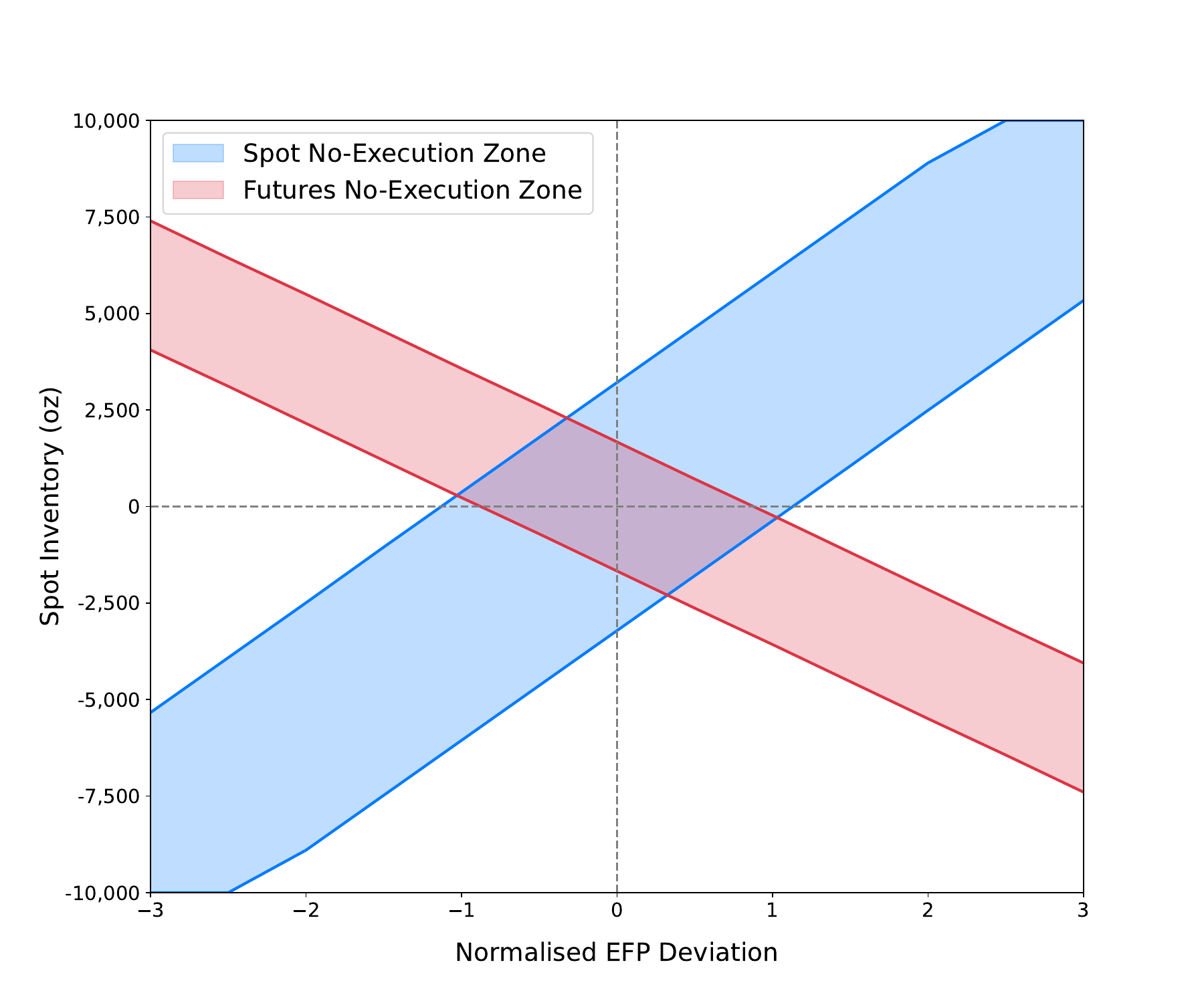}\\
\caption{Gold spot and futures no-execution zones as functions of spot inventory and volatility-normalised EFP price deviation.
Zero futures inventory, $\gamma=10^{-3}$ (left) and $10^{-4}$ (right).
No execution in shaded areas, selling the corresponding instrument above the upper boundary and buying below the lower boundary.
}
\label{onset}
\end{figure}

\begin{figure}[h!]
\centering
\includegraphics[width=0.49\textwidth]{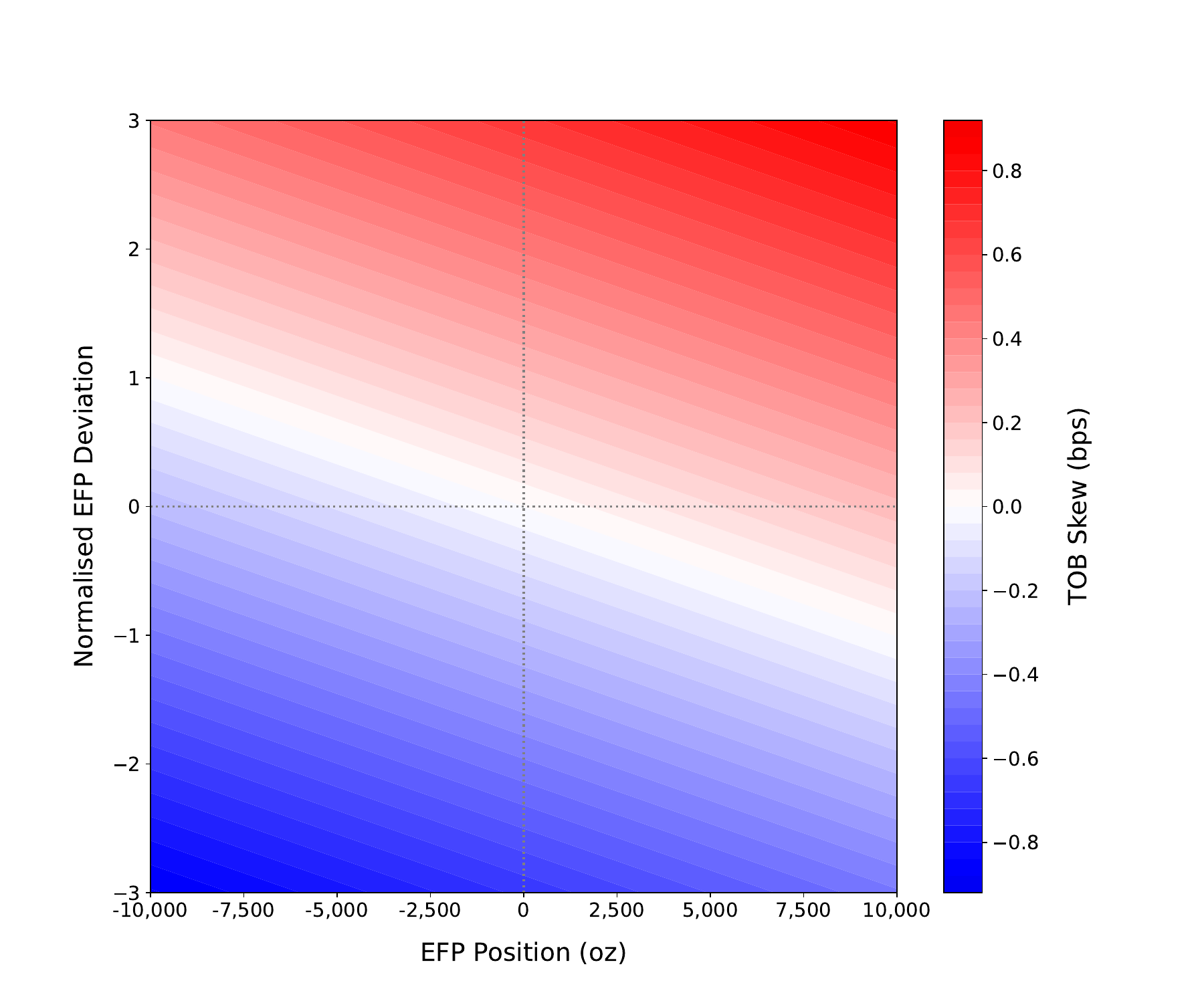}
\includegraphics[width=0.49\textwidth]{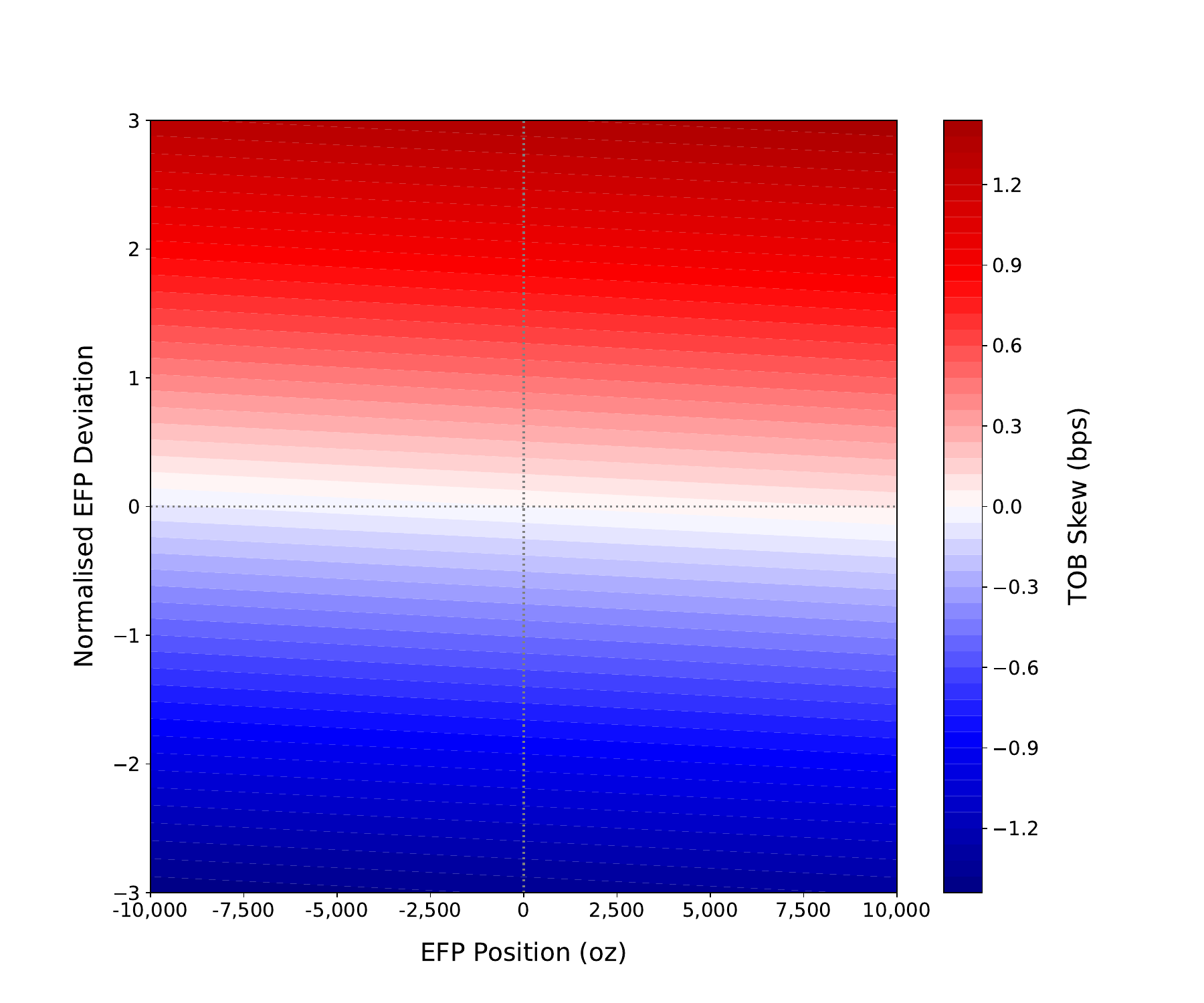}\\
\caption{Gold top of book spot skew as a function of EFP position and volatility-normalised EFP price deviation.
 $\gamma=10^{-3}$ (left) and $10^{-4}$ (right).
}
\label{skew}
\end{figure}

EFP mean reversion influences optimal controls. EFP deviation from the expected mean shifts the equilibrium inventory along with the corresponding execution onsets and quote skew, as shown in Figures~\ref{onset} and~\ref{skew}. Understandably, when the deviation is negative the dealer will tend to accumulate long EFP position, and vice versa. Figure~\ref{onset} demonstrates that EFP mean reversion will only lead to rare direct arbitrage opportunities under strong risk aversion. This is related to the cost of opportunistically entering into an EFP position (one would have to cross two spreads). The dealer will instead skew quotes in the required direction (see Figure~\ref{skew}) and wait for the opportunity to materialize while making spread. With a lower risk aversion, the appetite to capitalize on EFP mean reversion increases leading to opportunistic execution at extreme deviations (where the upper futures execution onset is below the lower spot execution onset and, similarly, where the lower futures execution onset is above the upper spot execution onset). The skew will be repurposed from EFP risk management to opportunistic skew on EFP deviation, as shown in Figure~\ref{skew}.\\

\begin{figure}[h]
\centering
\includegraphics[width=0.73\textwidth]{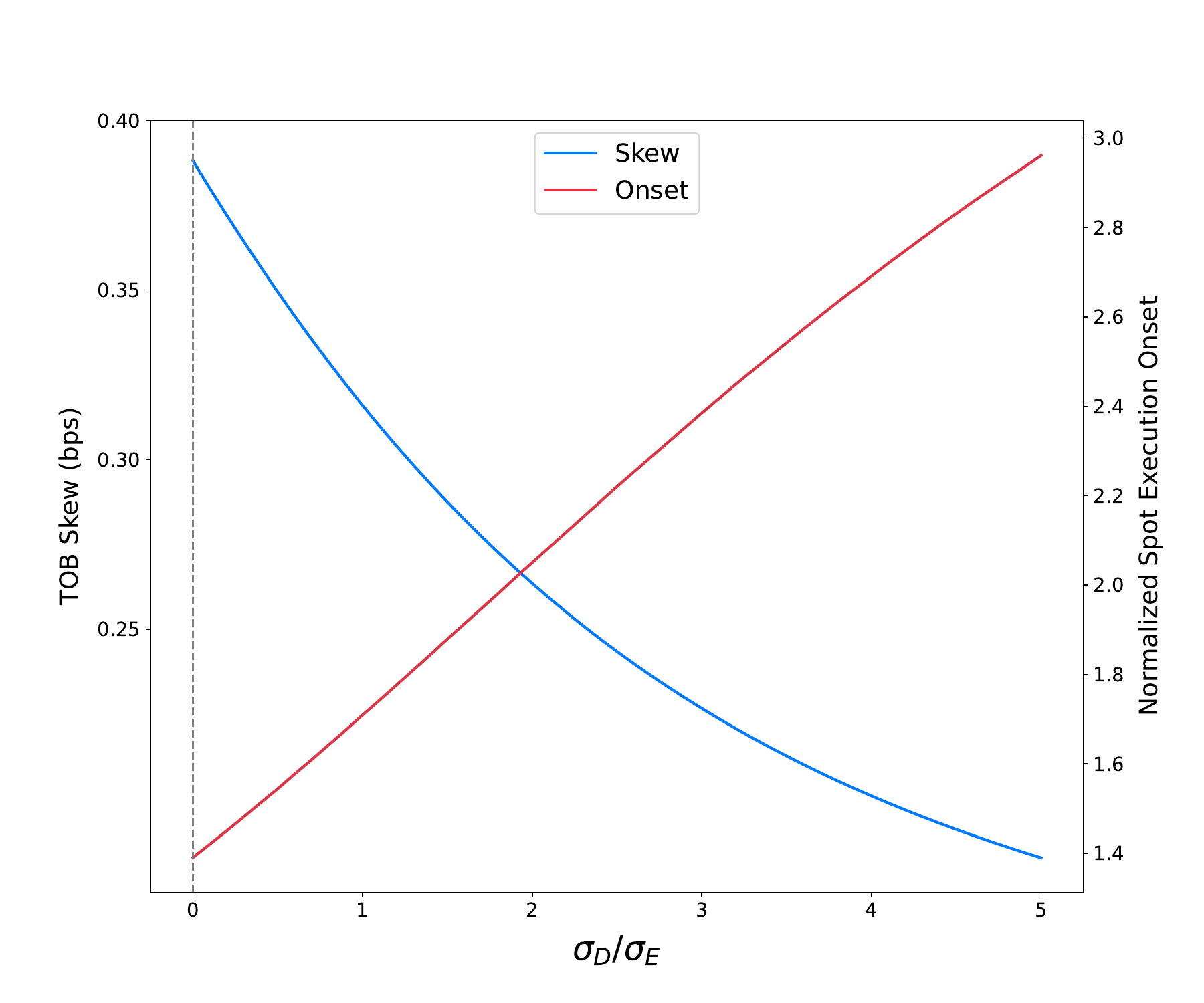}\\
\caption{Top of book skew for the volatility-normalized EFP deviation of $\epsilon \equiv E/\sigma_E = 1$ and $D=0$ along with the value of $\epsilon$ corresponding to spot execution onset as functions of normalized volatility of the mean, $\sigma_D/\sigma_E$, for zero spot and futures inventory and $D=0$. 
}
\label{nested}
\end{figure}

We introduced nested OU processes in this paper to mitigate the latter repurposing which is mainly due to overconfidence when using the OU model. Uncertainty on mean EFP deviation through the introduction of the stochastic process $(D_t)_{t\ge 0}$ creates additional risk that is taken into account through risk aversion, resulting in the dealer being less inclined to keep EFP position. Figure~\ref{nested} demonstrates the effect of $\sigma_D$ on the propensity to capitalize on EFP mean reversion. Here $k_D  = 0.2\,\text{day}^{-1}$, $\gamma = 3\cdot 10^{-4}$, and other parameters are as defined earlier. As expected, the larger is $\sigma_D$, the smaller is the skew and the higher is the arbitrage onset.\\ 

\begin{figure}[h!]
\centering
\includegraphics[width=0.65\textwidth]{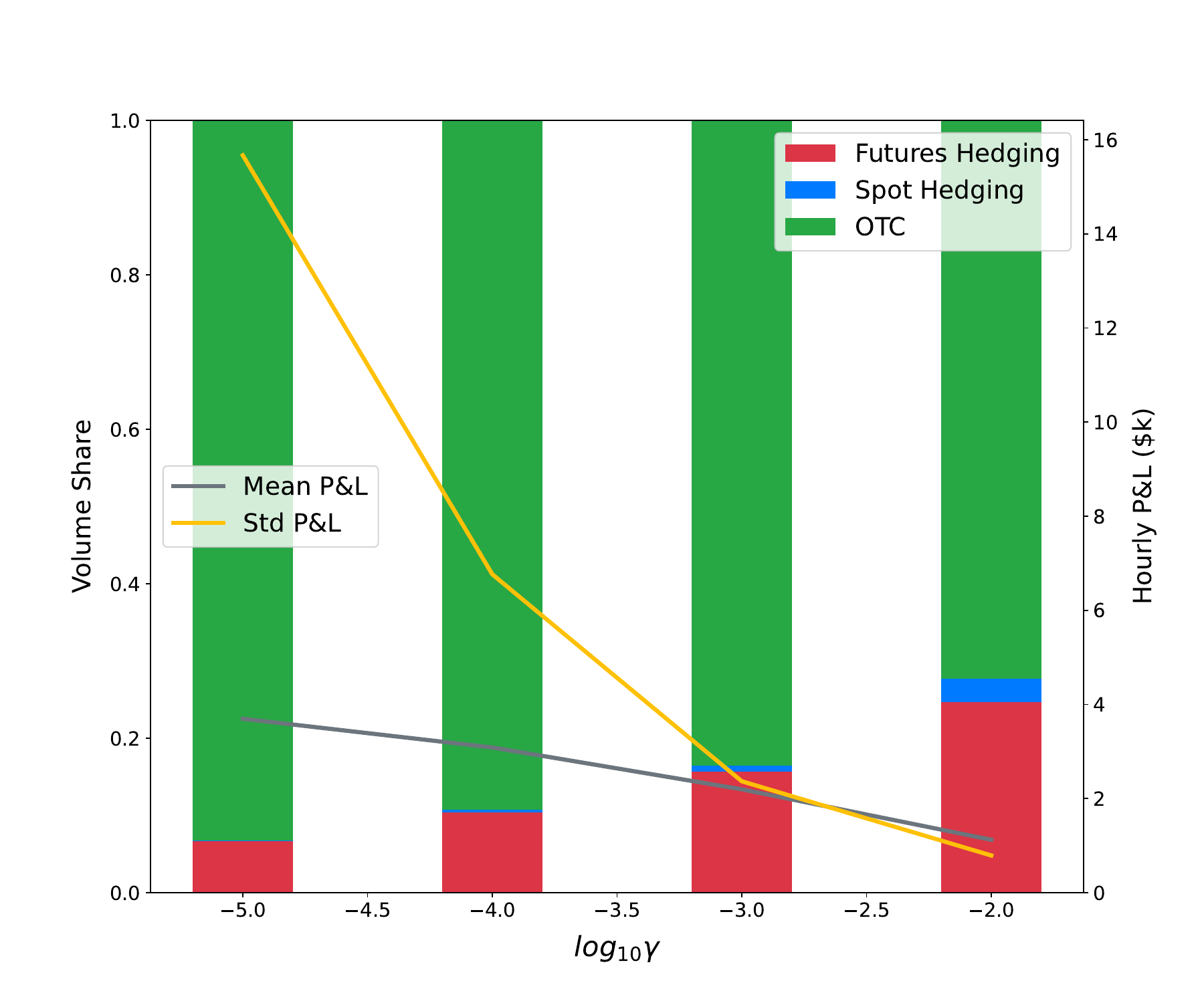}\\
\caption{Volume share of client trades and hedging, hourly P\&L and standard deviation of hourly P\&L as functions of risk aversion.
}
\label{volume_share}
\end{figure}

The choice of risk aversion is ultimately in the hands of the dealer. Figure~\ref{volume_share} demonstrates the effect of risk aversion on volume share of hedging, P\&L and risk. In this specific example, it is clear that decreasing $\gamma$ below $10^{-4}$ is questionable as the expected risk increases much faster than P\&L. Similarly, increasing $\gamma$ above $10^{-3}$ is not efficient.\\

\begin{figure}[h!]
\centering
\includegraphics[width=0.92\textwidth]{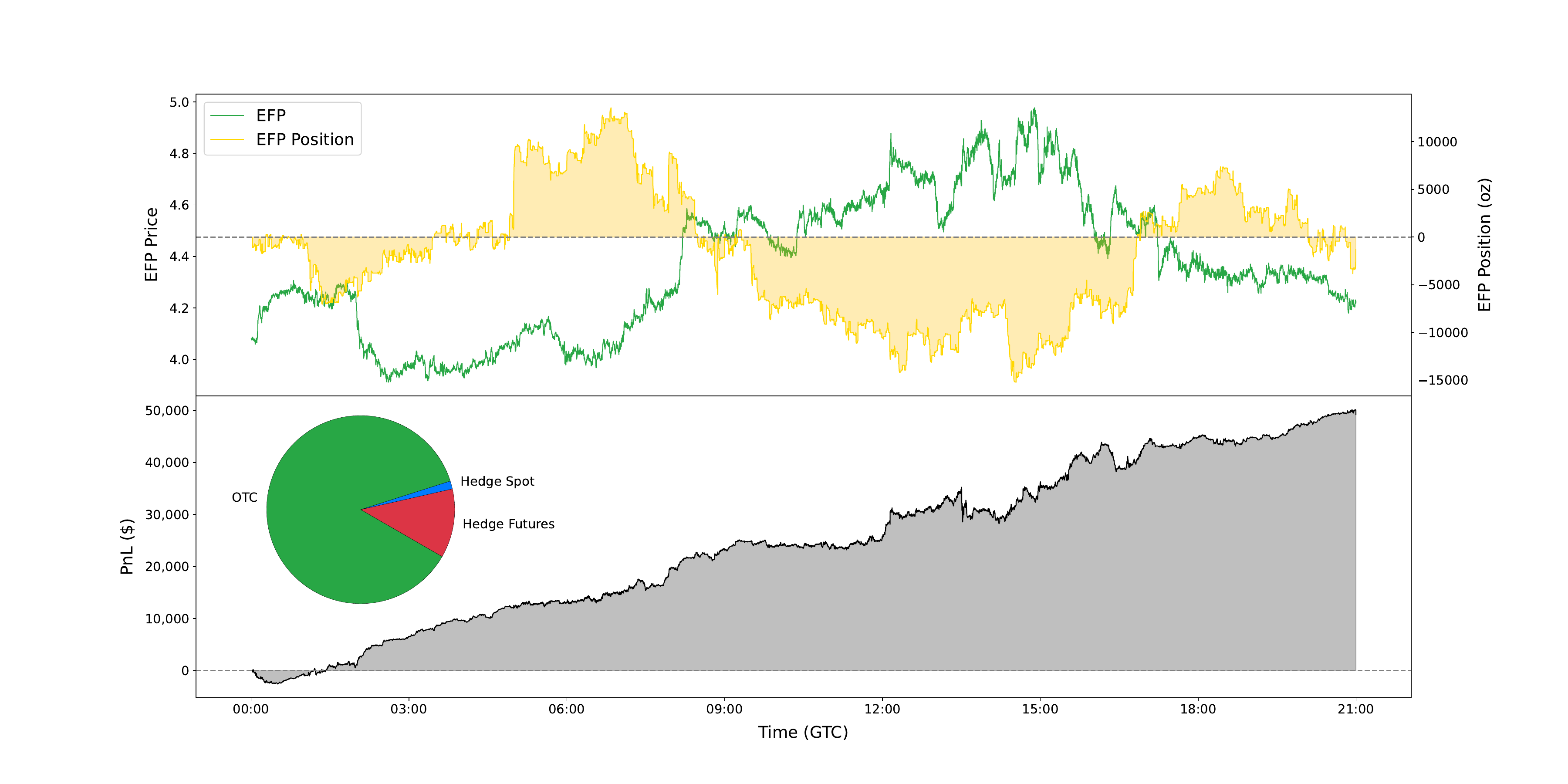}\\
\caption{Optimal market making strategy performance on real market data (XAUUSD, 12 January 2024) with simulated client flow and risk aversion of $\gamma = 3\cdot 10^{-4}$. Top chart displays implied EFP spread and dealer's EFP position during the day, bottom chart shows realised cumulative P\&L with open positions marked to market mid. Insert illustrates volume share of OTC and hedging trades.}
\label{trading_simulation}
\end{figure}

\begin{figure}[h!]
\centering
\includegraphics[width=0.49\textwidth]{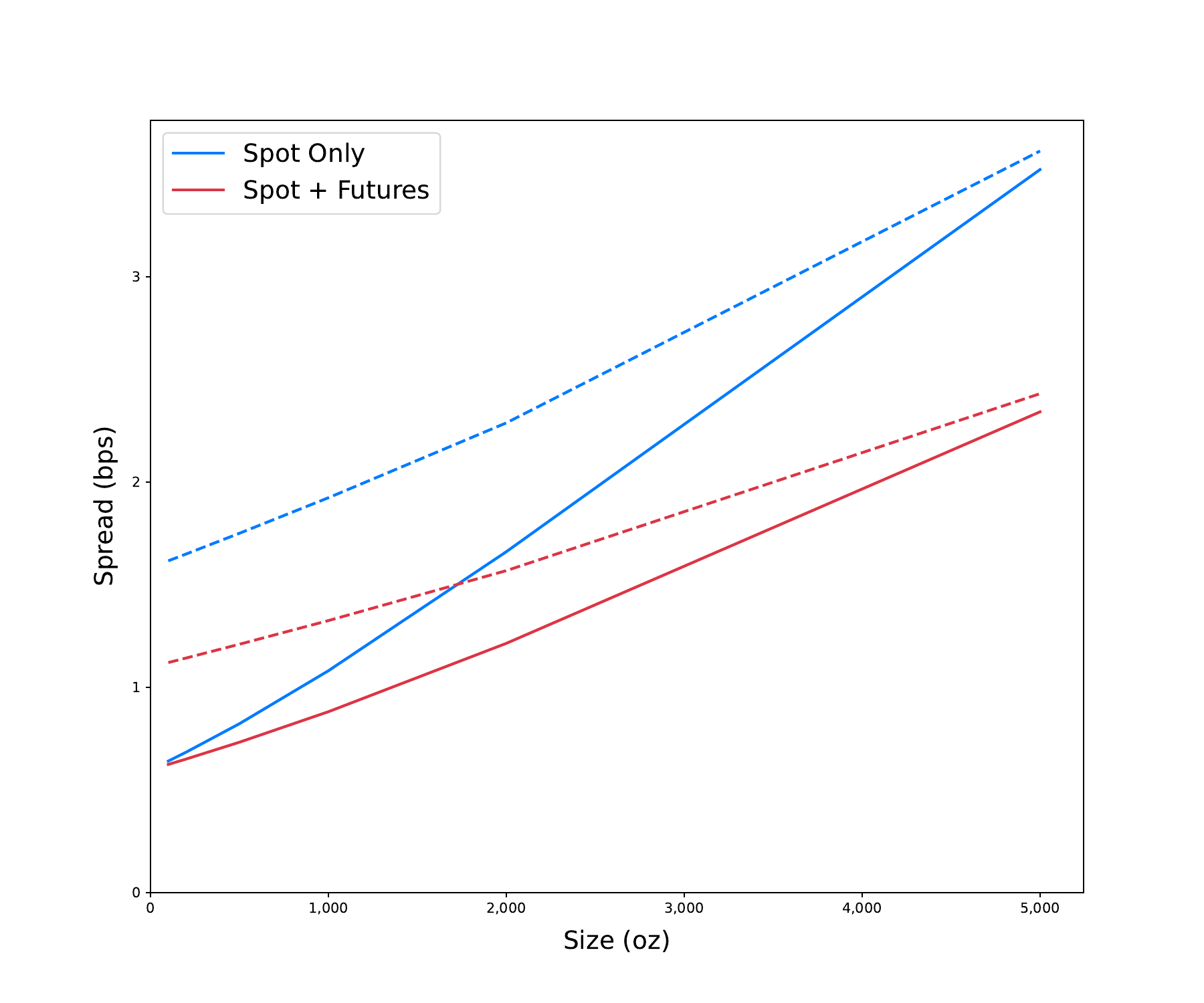}
\includegraphics[width=0.49\textwidth]{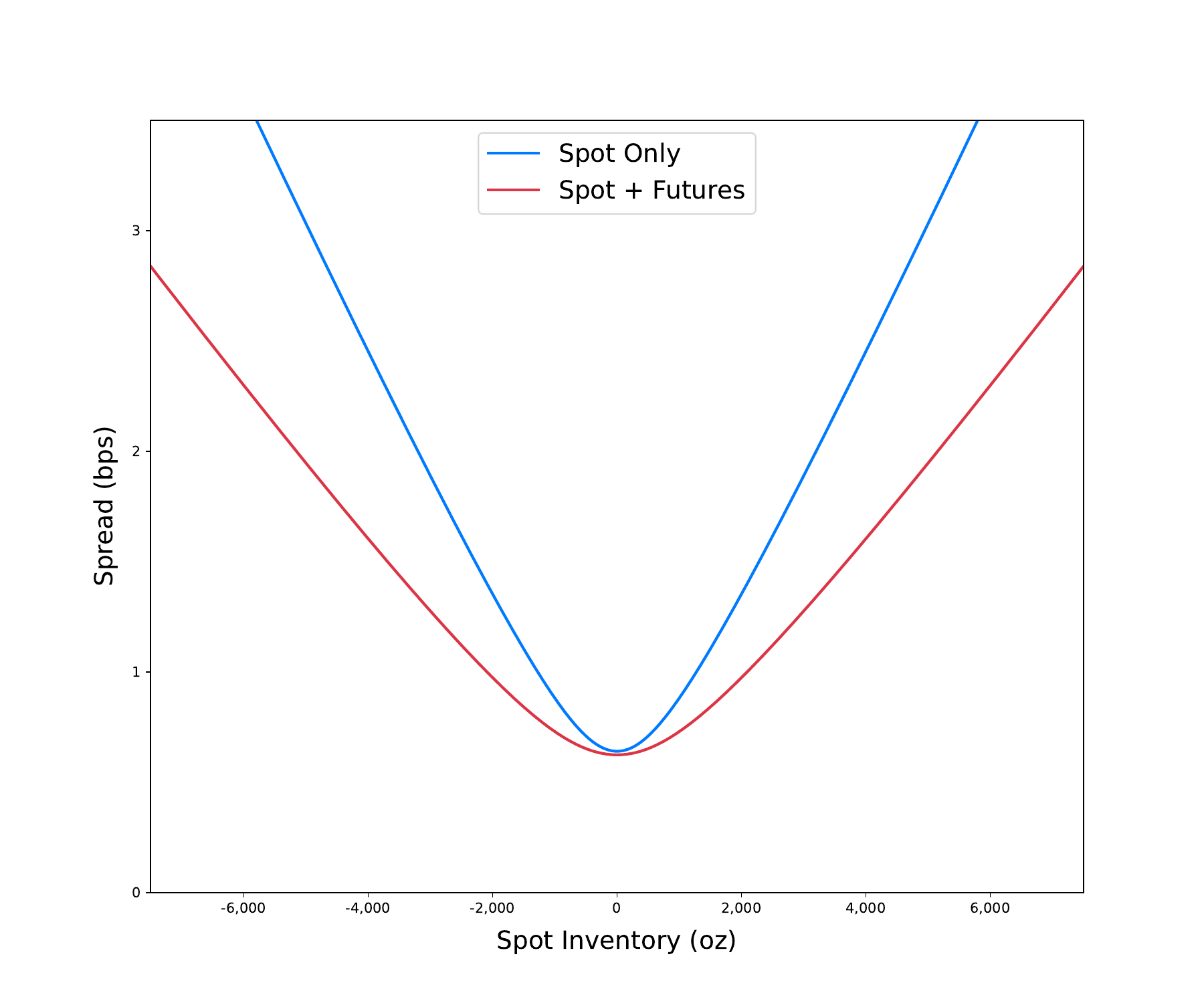}\\
\caption{Left: quoted bid-ask spread as function of size for spot inventories $q^S = 0$ (solid lines) and 2500 oz (dashed lines). Right: quoted top of book (100 oz) spread as a function of spot inventory. Hedging with spot only (blue) or with spot and futures (red). Risk aversion of $\gamma = 10^{-3}$, other parameters in the text.}
\label{futures}
\end{figure}

Figure~\ref{trading_simulation} illustrates the performance of the optimal strategy over a single trading day on real market data but with simulated client flow (uniform intensity throughout the day). EFP position clearly echoes the mean-reverting nature of EFP -- the dealer aims to keep EFP risk against the direction of EFP deviation. Volume share of futures execution significantly exceeds that of spot due to lower cost. Introducing futures hedging into risk management of OTC spot improves Sharpe ratio by at least 30\% (not shown), despite the additional round-trip cost of futures trading. It also allows to provide better prices to liquidity takers, as shown in Figure \ref{futures} where we compared the quoted bid-ask spreads in the absence and presence of a futures market.

\section*{Concluding Remarks}

We have extended a stochastic optimal control framework for OTC market making by incorporating co-integrated liquidity for hedging. Using a computationally efficient approximation technique, our methodology facilitates strategy optimization on demand in near real-time, potentially benefiting both electronic and voice traders. A specific example of spot gold market is analyzed in detail demonstrating efficient risk management benefiting from access to the significantly more liquid gold futures with reduced transaction costs while capitalizing on EFP mean reversion. In the spirit of the two-factor Hull-White model for interest rates, EFP spread is modeled by a nested Ornstein-Uhlenbeck process describing the observed multiple modes of relaxation corresponding to the diverse trading horizons of market participants. Interestingly, nested OU processes also appear as a way to robustify pure OU strategies with uncertainty in the mean reversion parameters, a commonly encountered case. They could find application in generalizing most of the optimization problems involving classical OU processes (see for instance the trading problem of \cite{lipton2020closed}).

\section*{Statement and acknowledgement}

The results presented in this paper are part of the research works carried out within the HSBC Research Initiative. The views expressed are those of the authors and do not necessarily reflect the views or the practices at HSBC. The authors are grateful to Richard Anthony, James Donaldson and Manuel Abellan-Lopez (HSBC) for helpful discussions and support throughout the project.

\bibliographystyle{plain}

\end{document}